\documentclass[preprint,12pt]{elsarticle}

\usepackage{amssymb}

\usepackage{hyperref}
\usepackage{siunitx}
\usepackage[version=4]{mhchem}
\usepackage{float}
\usepackage{textcomp}

\usepackage{tikz}
\usetikzlibrary{shapes,arrows, positioning}

\tikzstyle{decision} = [diamond, draw,
text width=4.5em, text badly centered, node distance=3cm, inner sep=0pt]
\tikzstyle{block} = [rectangle, draw,
text width=6.2em, text centered, rounded corners, minimum height=4em]
\tikzstyle{rdash} = [rectangle, draw, dashed,
text width=6.2em, text centered, rounded corners, minimum height=4em]
\tikzstyle{line} = [draw, -latex']
\tikzstyle{cloud} = [draw, ellipse, node distance=3cm,
minimum height=3em]

\usepackage{listings}
\usepackage{xcolor}

\definecolor{codegreen}{rgb}{0,0.6,0}
\definecolor{codegray}{rgb}{0.5,0.5,0.5}
\definecolor{codepurple}{rgb}{0.58,0,0.82}
\definecolor{backcolour}{rgb}{0.97,0.97,0.97}

\lstdefinestyle{mystyle}{
  backgroundcolor=\color{backcolour},   
  commentstyle=\color{codegreen},
  keywordstyle=\color{magenta},
  numberstyle=\tiny\color{codegray},
  stringstyle=\color{codepurple},
  basicstyle=\ttfamily\footnotesize,
  breakatwhitespace=false,         
  breaklines=true,                 
  captionpos=b,                    
  keepspaces=true,                 
  numbers=left,                    
  numbersep=5pt,                  
  showspaces=false,                
  showstringspaces=false,
  showtabs=false,                  
  tabsize=2
}

\usepackage{caption}

\usepackage[large,bf]{subfigure}

\newcommand{\uh}{\_\discretionary{-}{}{}}
\newcommand{\PC}[1]{\ensuremath{\left(#1\right)}}
\newcommand{\PD}[1]{\ensuremath{\left[#1\right]}}
\newcommand{\PE}[1]{\ensuremath{ \left\lbrace #1\right\rbrace}}

\newcounter{bla}

\journal{-}

\begin{document}

\begin{frontmatter}



\title{\textit{PyEquIon}: A Python Package For Automatic Speciation Calculations of Aqueous Electrolyte Solutions}


\author[eq]{Caio Felippe Curitiba Marcellos \corref{cor1}}
\author[eq]{Gerson Francisco da Silva Junior}
\author[eq]{Elvis do Amaral Soares}
\author[mat]{Fabio Ramos}
\author[eq]{Amaro G. Barreto Jr}

\cortext[author] {Corresponding author.\\\textit{E-mail address:} caiocuritiba@gmail.com}

\address[eq]{Department of Chemical Engineering, Center of Technology, Federal University of Rio de Janeiro\unskip, Rio de Janeiro\unskip, 21941-909\unskip, RJ\unskip, Brazil}
\address[mat]{Department of Applied Mathematics, Institute of Mathematics, Federal University of Rio de Janeiro, Brazil}

\begin{abstract}
In several industrial applications, such as crystallization, pollution control, and flow assurance, an accurate understanding of the aqueous electrolyte solutions  is crucial. Electrolyte equilibrium calculation contributes with the design and optimization of processes by providing important information, such as species concentration, solution pH and potential for solid formation. In this work, a pure Python library distributed under BSD-3 license was developed for the calculation of aqueous electrolyte equilibrium. The package takes as inputs the feed components of a given solution, and it automatically identifies its composing ions and the chemical reactions involved to calculate equilibrium conditions. Moreover, there is no established electrolyte activity coefficient model for a broad range of operational conditions. Hence, in this package,  built-in activity coefficient models are structured in a modular approach, so that the non-ideality calculation can be performed by a user provided function, which allows further research in the topic. The package can be used by researchers to readily identify the equilibrium reactions and possible solid phases in a user friendly language.
\end{abstract}


\end{frontmatter}



{\bf PROGRAM SUMMARY}

\begin{small}
\noindent
{\em Program Title:} \textit{PyEquIon}                                          \\
{\em CPC Library link to program files:} (to be added by Technical Editor) \\
{\em Developer's repository link:} \url{https://github.com/caiofcm/pyequion} \\
{\em Code Ocean capsule:} (to be added by Technical Editor)     \\
{\em Licensing provisions(please choose one):} BSD 3-clause     \\
{\em Programming language:} Python                              \\
{\em Nature of problem:}
Calculation of aqueous electrolyte equilibrium in a pure python package with code exportation for embedding in dynamic simulations. 
\\
{\em Solution method:} 
A recursive algorithm identifies the system reactions and chemical species. Possible candidates of solid phases are also identified. The equilibrium equations are defined based on a database of reactions and associating the involved species. The nonlinear system is composed of the equilibrium equations, charge balance, mass balances and optionally gaseous carbon dioxide equilibrium or pH. Activity coefficient models, such as Debye-H\"{u}ckel and Pitzer, are used based on a database containing its parameters. The nonlinear system is solved with a Newton-Raphson method. The software allows automatic jacobian generation and code exportation for incorporation in differential algebraic equations.
\\


\end{small}

\section{Introduction}

Aqueous electrolyte solutions are of major interest in many industrial and environmental applications and comprise different research fields, such as geological, pharmaceutical, biological, chemical, nuclear and petroleum engineering. Equilibrium calculations of aqueous electrolytes systems are often implemented in mathematical models for designing and monitoring of processes \cite{curitibainferring,takano2000computer}. However, despite the widespread literature on electrolyte systems, the theory of electrolyte thermodynamics is not completely understood with a lack of consensus in model formulations \cite{prausnitz1998molecular}. Some of the main challenges in the thermodynamic modeling of electrolyte solutions are \cite{may2017thermodynamic}: (i) the choice of computer code and appropriate parameter database; (ii) proper implementation of published models, because the informed formulation can be often inadequate; and (iii) an inappropriate emphasis of studies in modifications and extension of models without substantial physical foundation.

There are reliable open-source and commercial computational tools for calculating chemical equilibrium of aqueous solution with a comprehensive set of compounds and thermodynamic database, \cite{may_jess_1991, haase_uncertainty_2013, parkhurst2013description,rice2009scalesoftpitzer, kaasa1998multiscale}.
Some of them were developed few decades ago in technologies that are now outdated, so that it  an be cumbersome to set the computational environment for its proper usage. Additionally, for some packages, it can be difficult for a non-programmer to access the code and to modify it for a specific need. For non-expert researchers or industrial analysts, chemical electrolyte packages are often used in a black-box manner, which can make studies less reproducible and error prone.

General-purpose open source computing languages, such as Python, have garnered great attention from the scientific community due to dynamic prototyping and high-level data structure, which allows for both sophisticated development and scripting \cite{kirsanskas_qmeq_2017}. As a multipurpose language, it enables interactions between services enhancing the overall research workflow, such as data manipulation, model creation, and data visualization. Moreover, it provides good code readability, interoperability with other languages, a built-in documentation system, rich collection of data structures and a variety of libraries for scientific computing \cite{koepke2011python, mortensen_high_2016}.


In this work, we present a computational tool for the aqueous speciation calculation with automatic equilibrium reactions determination. Given a set of input components, the software enables the automatic ions identification and the involved equilibrium reactions. The final equation system is formed by the electroneutrality equation, mass balances and, if necessary, a closing equation. For the efficient usage in iterative numerical methods, the package offers just in time compilation, automatic jacobian generation and code exportation of a residual function for incorporation in differential algebraic equation systems. A modular approach for nonideality calculation allows the incorporation of user-provided functions for the species activity coefficient. This tool is a contribution for the researchers' understanding of electrolyte chemical reactions, and it can also be used for numerical simulation.

\section{Aqueous Electrolyte Equilibrium Overview}

A mixture of solutes in a solvent can originate various species, such as free ions, undissociated molecules, ion pairs, complexes or chelates, and micelle cluster \cite{wright2007introduction}. Thus, it is necessary to identify all possible species present in the solution, and all the equilibrium reactions between them. 

The fundamental law of mass action \cite{massaction} states that for a given reaction

\begin{equation}
  \ce{a A + b B <=> c C + d D},
  \label{eq:reac}
\end{equation}

the reaction equilibrium constant $K$ can be defined as:

\begin{equation}
  K = \frac{\left\{ C \right\}^{c}\left\{ D \right\}^{d}}{\left\{ A \right\}^{a}\left\{ B \right\}^{b}},
\end{equation}

\noindent with the bracket operator denoting activity.

The activity of a compound $\PE{A}$ is the effective concentration of $A$ in the solution given by $\PE{A} = \gamma_A \PD{A}$, where $\PD{A}$ is the molal concentration of $A$ and $\gamma_A$ is its activity coefficient. The activity coefficient describes the deviation from the ideal solution, and the nonideality is primarily due to electrostatic interactions between the ions. It can be observed experimentally, for instance, if one computes the equilibrium constant directly with ions concentration for various ionic strength the equilibrium constant may vary. Another observation is the variation of molar conductivities of strong electrolytes within a concentration range \cite{wright2007introduction}.


A common approach for calculating activity coefficients is to use the Debye-H\"{u}ckel (DH) theory, which was derived from the Poisson-Boltzmann equation considering as the source of nonideality the electrostatic double-layer interaction \cite{debye1923theorie}. The Debye-Hückel limiting law, for very dilute solutions, relates the activity coefficient with the ionic strength (Eq. \ref{eq_I}) and is given by Eq. \ref{eq_gamma}:

\begin{equation}
  I = \sum_{i\in C} \PD{i}{z_i}^2,
  \label{eq_I}
\end{equation}

\begin{equation}
  \log\gamma_{A} = - A_{\phi} z_{i}^{2} \sqrt{I},
  \label{eq_gamma}
\end{equation}

\noindent where $z_{i}$ is the charge of specie, $C$ is the set of all species and $A_{\phi}$ a temperature dependent parameter given by Eq. \ref{eq_Aphi}:

\begin{equation}
  A = \left( \frac{e^{2}}{\epsilon_{0}\ \epsilon_{r}\text{RT\ }} \right)^{\frac{3}{2}}\frac{N_{A}}{8\pi}\sqrt{2\rho_{w}}
  \label{eq_Aphi}
\end{equation}

\noindent with $e$ as the electron charge, $\epsilon_{0}$ and $\epsilon_{r}$ are the vacuum and relative permittivity, respectively, $\rho_{w}$ is the solvent density, $N_{A}$ is the Avogadro's constant and $T$ is the temperature \cite{prausnitz1998molecular}.

The DH model is only accurate for very dilute solutions because it does not take into account some factors, for example, short-range interaction forces, which are very important for more concentrated solutions. For concentrated solution the main approaches are \cite{prausnitz1998molecular}: (i) Physical models: are derivation from the DH model in which the most cited is the Pitzer model and consider interactions due to excluded volume and Van der Waals attraction; (ii) Chemical models: the nonideality is attributed to semi-stable chemical species (solvated ions), for instance, the Robinson and Stokes model and (iii) Local-composition models: the physical van der Waals interaction are considered as a function of local composition, such as NRTL, Wilson or UNIQUAC models. All those approaches require the fitting of parameters from experimental data.

From the nonideality models derived from DH equation, there are \cite{zemaitis2010handbook, pitzer1993thermodynamics, appelo2004geochemistry}: (i) the extended DH (1923), (ii) the Guggenheim Model (1935) with includes two terms for ion interaction and is a precursor of the Pitzer model; (iii) the Davies model (1938) with received attention since has no ion-specific parameter; (iv) H\"{u}ckel Equation (or B-dot) that is used in this work and has two ion-specific parameters; (v) a polynomial DH model by the National Bureau of Standards \cite{goldberg1981evaluated} for higher ionic strength; (vi) corrected DH (1983) with the account for finite ion size; (vii) the Bromley \cite{bromley1973thermodynamic} activity model that can model multicomponent system of $I$ up to 6 molal and (viii) the Pitzer Equation which is a semi-empirical extended DH with virial expansion and requires high-order interaction parameters for high concentrated solutions.

A fully experimental nonideality method is the Meissner Reduced Activity Coefficient Model \cite{meissner1972activity} with regression for a modified activity variable related to the ionic strength. Local-composition models are, for instance, the Multicomponent e-UNIQUAC \cite{chen1986local} and the e-NRTL (Cardoso et. al., 1986). Recently, the Generalized DH model \cite{liu2019generalized} was proposed based on Poisson-Fermi equation with the goal of describing the activity coefficient with fewer parameters, although it still requires more validation for multicomponent systems. 



The Davies equation, Eq. \ref{eq_davies_model}, is often used for low concentration solutions (up to 0.5 M). The B-dot equation, Eq. \ref{eq_dh_model}, is a virial-type expansion for the Debye-H\"{u}ckel equation where $b$ is an adjustable parameter, which is the difference between the observed activity coefficient logarithm and the calculated activity coefficient logarithm by the Debye-H\"{u}ckel equation. With this approach, this model becomes accurate at concentrations up to 1 M \citep{zemaitis2010handbook}.

\begin{equation}
  \log\gamma_i = - A z_{i}^{2}\frac{\sqrt{I}}{1 + \sqrt{I}} - 0.3I.
  \label{eq_davies_model}
\end{equation}

\begin{equation}
  \log\gamma_i = - \frac{z_{i}^2 A \sqrt{I}}{1 + B a_{i}\sqrt{I}} + b_iI.
  \label{eq_dh_model}
\end{equation}

For more concentrated solutions, the most used model is the ion-interaction model proposed by Pitzer \cite{pitzer_thermodynamics_1973}. This model takes into account long and short-range interactions, where long-range forces are presented by the extended DH model, and short-range forces are presented as a virial type expansion with parameters of interaction between two or three particles of the solute \cite{prausnitz1998molecular, zemaitis2010handbook, pitzer1993thermodynamics, pabalanandpitzer1987, pitzer1975thermo}. The parametric expressions used by Pabalan and Pitzer \cite{pabalanandpitzer1987} are presented in Eq. \ref{eq_pitzer_cation} and in Eq. \ref{eq_pitzer_anion} for an M cation and an X anion of interest in a mixture of electrolytes.

\begin{equation}
   \begin{array}{cl}
      \ln \gamma_{M} = z_{M}^{2}F + \sum_{a} m_{a'} (2B_{Ma} + ZC_{Ma} + \sum_{c} m_{c}(2\Phi_{Mc} + \sum_{a} m_{a}\psi_{Mca})\\
      +\sum_{a} \sum_{a'} m_{a} m_{a'}\Psi_{Maa'} + |z_{M}|\sum_{c}\sum_{a} m_{c} m_{a}C_{ca} + \sum_{n} m_n (2\lambda_{nM})
   \end{array}
   \label{eq_pitzer_cation}
\end{equation}

\begin{equation}
  \begin{array}{cl}
      \ln \gamma_{X} = z_{X}^{2}F + \sum_{c} m_{c} (2B_{cX} + ZC_{cX} + \sum_{a} m_{a}(2\Phi_{Xa} + \sum_{c} m_{c}\psi_{Xac})\\
      +\sum_{c} \sum_{c'} m_{c} m_{c'}\Psi_{Xcc'} + |z_{X}|\sum_{c}\sum_{a} m_{c} m_{a}C_{ca} + \sum_{n} m_n (2\lambda_{nX})
   \end{array}
  \label{eq_pitzer_anion}
\end{equation}

Where $F$ account the long-range interaction forces and its value is calculated according to Eq. \ref{par_F} and $Z$ is defined by Eq. \ref{par_Z}, $c$ and $c'$ are cations and $a$ and $a'$ are anions, $m_{c}$ and $m_{a}$ are the concentrations on the cation and anion molal scale, respectively.

\begin{equation}
    \begin{split}
        F = A_{\phi} \left[ \frac{\sqrt{I}}{1+b\sqrt{I}} + \frac{2}{b} \ln(1+b\sqrt{I}) \right] + \sum_{c} \sum_{a} m_c m_a B'_{ca} + \\
        \sum_{c} \sum_{c'} m_c m_{c'} \Phi'_{cc'} +  \sum_{a} \sum_{a'} m_a m_{a'} \Phi'_{aa'}
    \end{split}
  \label{par_F}
\end{equation}

\begin{equation}
  Z = \sum_{i} m_{i}\left| z_{i} \right|
  \label{par_Z}
\end{equation}

The $B$ and $B'$ parameters are calculated by Eq. \ref{par_B1} and \ref{par_B2}. The parameters $\beta$ are determined empirically from binary systems. The variable $B'$ is the ionic strength derivatives of $B$. The $C$ term is determined by Eq. \ref{par_C}, where $C^{\phi}$ parameter is also determined empirically from binary data.

\begin{equation}
B_{MX} = \beta_{MX}^{(0)} + \beta_{MX}^{(1)}g(\alpha\sqrt{I}) + \beta_{MX}^{(2)}g(\alpha\sqrt{I})
\label{par_B1}
\end{equation}
\begin{equation}
B' = \frac{1}{I}(\beta_{MX}^{(1)}g'(\alpha\sqrt{I}) + \beta_{MX}^{(2)}g'(\alpha\sqrt{I}))
\label{par_B2}
\end{equation}
\begin{equation}
g(x) = \frac{2}{x^{2}}[1-(1+x)\exp(-x)]
\label{par_B3}
\end{equation}
\begin{equation}
g'(x) = \frac{-2}{x^{2}}[1-(1+x+\frac{x^{2}}{2})\exp(-x)]
\label{par_B4}
\end{equation}

\begin{equation}
C = \frac{C_{MX}^{\phi}}{2\sqrt{\left| z_{i}z_{j} \right|}}
\label{par_C}
\end{equation}  

On the other hand, the parameter $\psi$ can be obtained from experimental data of mixed electrolytes and the parameter $\Phi$ and $\Phi'$ are determined as described in Pitzer (1975) \cite{pitzer1975thermo}.

\section{Software description}

\textit{Pyequion} is a free python library distributed under BSD license for the automatic determination of equilibrium reaction and species. In short, for a user provided component mixture, reactions as in Eq. \ref{eq:reac} are determined together with the involved species. Reaction constants are obtained from a reaction database and the activity coefficients are calculated from a selected model with species parameters from a database. The nonlinear model can be exported for embedding in dynamic simulations.

\begin{figure}[H]
    \centering
    \resizebox{0.5\columnwidth}{!}{%
    \begin{tikzpicture}[node distance = 2cm, auto]
        \node [block] (init) {Initial Compounds and Settings};
        \node [block, below of=init, node distance = 2.5cm] (Reacs) {Reaction Identification};
        \node [rdash, right of=Reacs, node distance = 3.5cm] (DB-Reacs) {DB Reactions (Solution and Solid)};
        \node [block, below of=Reacs, node distance = 2.5cm] (Create) {Create Equilibrium Object};
        \node [rdash, left of=Create, node distance = 3.5cm] (DB-Species) {DB Species};
        \node [block, below of=Create, node distance = 2.5cm] (Setup) {Setup nonideality model};
        \node [cloud, left of=Setup, node distance = 3.5cm, align=center] (setup_fn) {\texttt{setup $\gamma$} \\ \texttt{function}};
        \node [block, below of=Setup, node distance = 2.5cm] (Solve) {Solve Equilibrium};
        \node [block, below of=Solve, node distance = 2.5cm] (Props) {Calculate Properties};
        \node [cloud, right of=Solve, node distance = 3.5cm, align=center] (calc_fn) {\texttt{calc $\gamma$} \\ \texttt{function}};

        \path [line] (init) -- (Reacs);
        \path [line, dashed] (DB-Reacs) (DB-Reacs.west) -- (Reacs.east) (Reacs);
        \path [line] (Reacs) -- (Create);
        \path [line, dashed] (DB-Species) |- (Create);
        \path [line] (Create) -- (Setup);
        \path [line, dashed] (setup_fn) -- (Setup);
        \path [line] (Setup) -- (Solve);
        \path [line, dashed] (calc_fn) -- (Solve);
        \path [line] (Solve) -- (Props);
    \end{tikzpicture}
    }
    \caption{Illustrative scheme for the required steps in a equilibrium calculation.}
    \label{fig_scheme}
\end{figure}
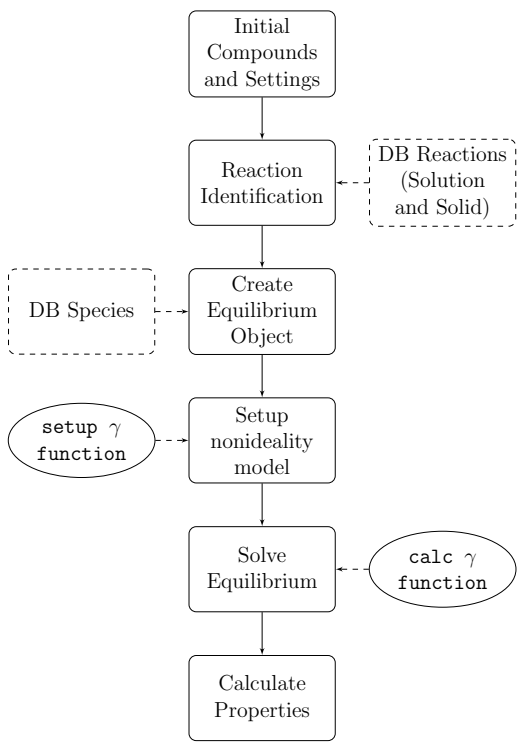

A representational scheme of the internal steps for equilibrium calculation is illustrated in Figure \ref{fig_scheme}. The identification of the reactions involved is the main part of the
computational tool. A database file containing various ionic equilibrium reactions in the JSON (Javascript Object Notation) or python dictionary format was created from the database file distributed with PHREEQC \cite{parkhurst2013description}. When all species in one side of an equilibrium reaction are known, the species on the other side can be formed. Thus, they are added to the system. Hence, by a recursive procedure, all reactions are traversed, checking for the condition of all species known in one side of the equilibrium reaction. The recursive algorithm is applied to all the new compounds added to the system.

At the end of the reaction identification procedure, the reactions and the present species are identified. The species are further processed to get information from its textual representation: its charge and the composing elements with their coefficients, by recognizing that each new element starts with an uppercase letter. This approach avoids redundancies in the species database.

Each chemical compound is related to another JSON (or python \textit{dict} datatype) database file of species with the required information for the nonideality model. The database file contains ion specific and interaction parameters that can be used by the chosen nonideality model. In this software version, the following models and companion parameters files are provided: (i) extended Debye-H\"{u}ckel equation or \textit{B-dot} \cite{plummer1976wateqf,parkhurst1990ion}; (ii) Bromley method with individual ions parameter \cite{bromley1973thermodynamic}; (iii) SIT formula \cite{grenthe_modelling_1997}; (iv) Pitzer method \cite{pitzer_thermodynamics_1973}. The parameters for the species activity coefficient calculation were obtained from the PHREEQC database files for \textit{B-dot}, SIT and Pitzer after conversion to the universally known structured file format (JSON). The parameters for the Bromley method were obtained from the original paper \cite{bromley1973thermodynamic} using the individual ion correlation. The parameters files are not exhaustive and interaction parameters may be missing. Users can provide new thermodynamic models using a setup function, which relates the species to the database, and the calculation function for the decimal logarithm of the species activity coefficient.



\subsection{Software Functionalities}

The following list enumerates the main functionalities provided by \textit{pyequion} v0.0.5.

\begin{itemize}
    \item Pure python package, without compiled dependencies, making it easy to install by non skilled programmers;
    \item Automatic identification of species and reactions for a given input compound list;
    \item Suitable integration with \textit{jupyter} notebooks for latex formatted display of reactions, which allows students and researchers to improve their understanding of the system;
    \item Built-in activity coefficient models: \textit{B-dot}, \textit{Bromley}, \textit{SIT} and \textit{Pitzer;}
    \item Modular approach for user defined functions for the calculation of activity coefficients;
    \item Reactions and parameters database already in convenient data interchange format (JSON);
    \item The \texttt{SolutionResult} class provide properties of the solution equilibrium, as: pH, conductivity, ionic strength, concentrations and activities, dissolved inorganic carbon, solid phases saturation index, ionic activity product or the concentrations in case of equilibrium with a solid phase and the reactions used;
    \item Solid phase equilibrium to calculate precipitated concentration and solubilities;
    \item Option for just-in-time compilation of the nonlinear residual calculation using the package \texttt{numba} \cite{10.1145/2833157.2833162};
    \item Code generation to export the residual calculation as a python function, hence allowing the calculation to be used in environments without the package and to be coupled in phenomenological models, such as in differential algebraic equations;
    \item Jacobian generation to a source file with the aid of symbolic computation (\texttt{sympy} \cite{meurer2017sympy}), which can also be just-in-time compiled;
    \item Henry law for \ce{CO2} vapour-liquid equilibrium with ideal or with the Peng-Robinson equation of state;
    \item Tutorial with relevant use cases.
\end{itemize}



\subsection{Software validation}

The software was validated against the output of the reference software PHREEQC \cite{parkhurst2013description} for low solute concentrations and in atmospheric conditions using the \texttt{pytest} framework for automatic unit tests. The software is accompanied by sample code comparing against experimental data for mean activity coefficients. The user should be aware that electrolyte equilibrium models are not established in the literature \cite{horbrand2018validation, haase2013uncertainty, may2017thermodynamic} and the results should be analyzed with caution, specially for modifications in conditions of temperature, pressure and concentrations.

\section{Illustrative Examples}

\subsection{Basic usage}

\begin{figure}[H]
    \centering
    \includegraphics[width=\linewidth]{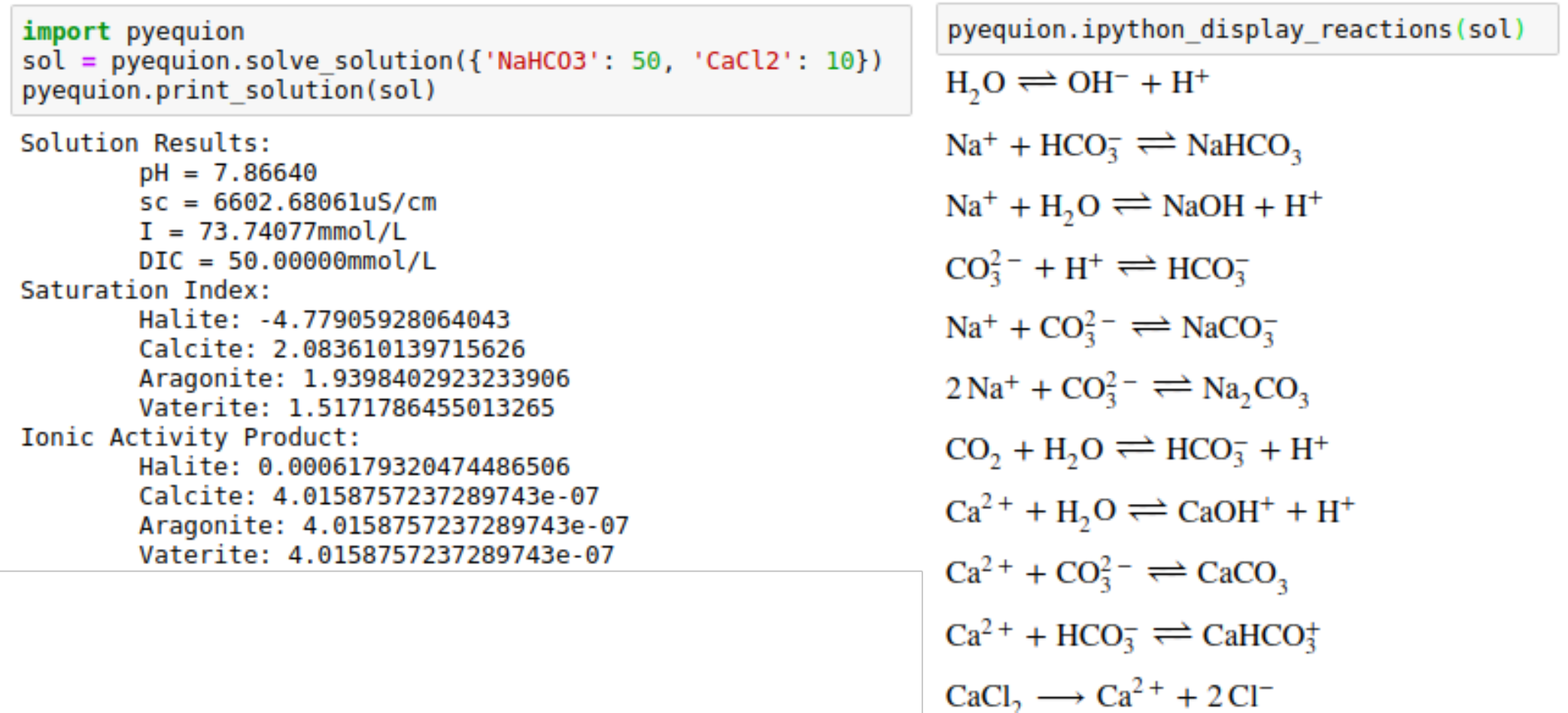}
    \caption{Example of equilibrium calculation for \ce{NaHCO3} and \ce{CaCl2} system using \textit{pyequion} showing the involved reactions.}
    \label{fig:nahco3-cacl2}
\end{figure}

Figure \ref{fig:nahco3-cacl2} depicts the \textit{pyequion} usage for aqueous equilibrium for the mixture of \ce{NaHCO3} and \ce{CaCl2}. The compositions are specified from the molarity of each component in millimolar and the closed system solution is considered by default. In this case, the calculated pH is \num{7.87} and the system is supersaturated for the calcite calcium carbonate polymorph, but not for halite. 

\begin{figure}[H]
    \centering
    \includegraphics[width=\linewidth]{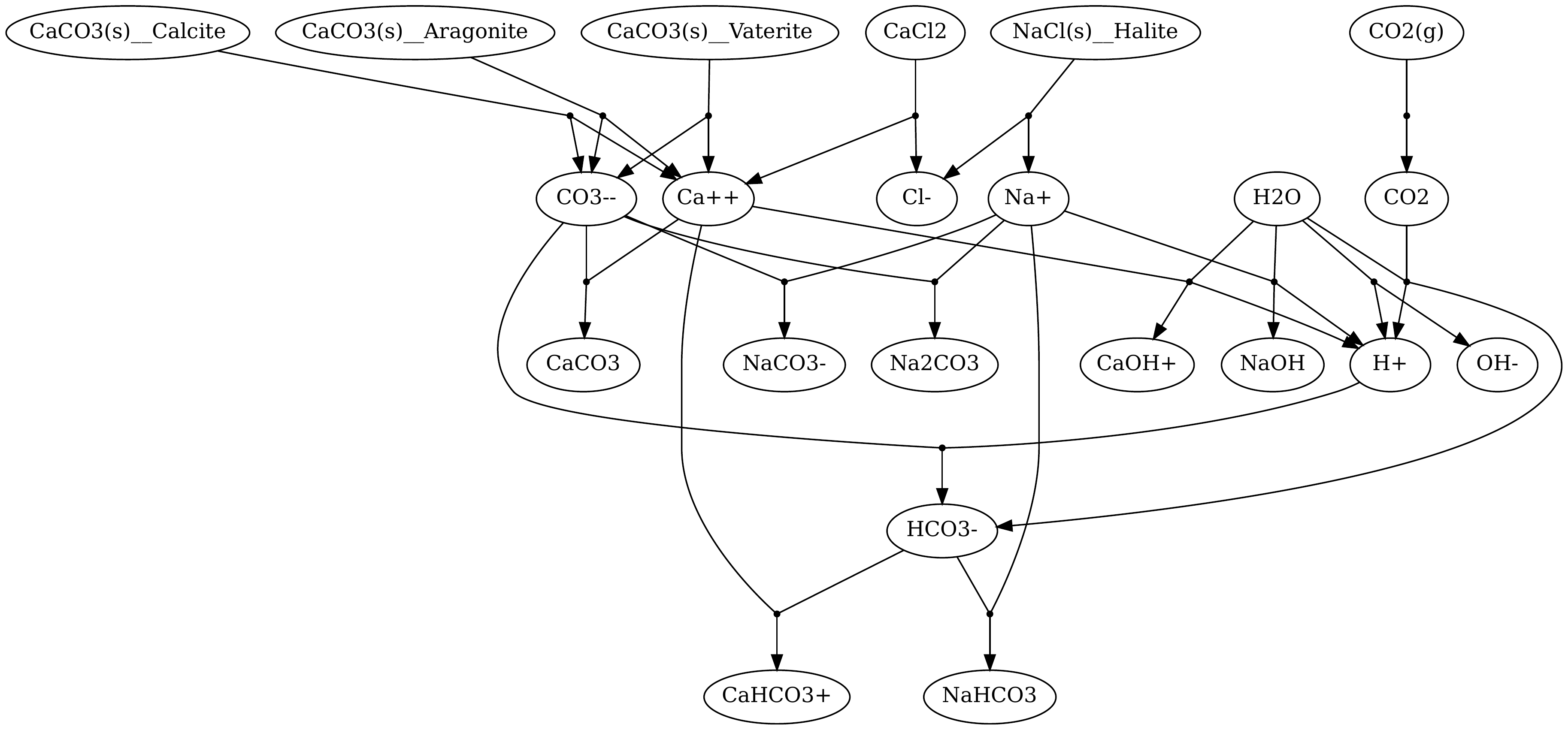}
    \caption{Graph of reactions for the \ce{NaHCO3} and \ce{CaCl2} open system with solid formation.}
    \label{fig:graph-reactions}
\end{figure}

There are other options available that are described in the package documentation and exemplified in Listing \ref{list_use1}. The example shows the solution for the same mixture modifying arguments as: (i) open system, allowing the equilibrium with calcite solid phase; (ii) selecting Peng-Robinson model for the fugacity coefficient in the vapour phase; (iii) adjusting the \ce{CO2} partial pressure for the atmospheric (default); and (iv) using one of the available activity coefficient model, the Pitzer method. Besides listing the involved reactions and species, the software can generate a graph of the reactions, which provides intuition on the species' connections, as shown in Fig. \ref{fig:graph-reactions}. 

\begin{lstlisting}[language=Python, label=list_use1, caption=Example of using \textit{PyEquIon} for aqueous equilibrium calculation modifying calculation options.]
    import pyequion
    sol = pyequion.solve_solution(
        {'NaHCO3': 50, 'CaCl2': 10},
        allow_precipitation=True,
        close_type=pyequion.ClosingEquationType.OPEN,
        activity_model_type='pitzer',
        solid_equilibrium_phases=['Calcite'],
        co2_partial_pressure=pyequion.pCO2_ref,
        fugacity_calculation='pr',
    )
    \label{list_use1}
\end{lstlisting}

\subsection{Calculating Mean Activity Coefficient with Pitzer Model}

In this case study, the aqueous electrolyte equilibrium of \ce{NaCl}, \ce{CaCl2} and \ce{MgCl2} are separately calculated using the Debye-H\"{u}ckel (\textit{B-dot}), SIT, Bromley with individual ion parameters and the Pitzer ion association method. The setup and calculation functions are provided by the package for the models and can be selected as presented in Listing \ref{lst-case-pt}.



\begin{lstlisting}[language=Python, label=lst-case-pt, caption=Calculating mean activity coefficient for \ce{CaCl2} with different thermodynamic models.]
sys_eq = pyequion.create_equilibrium(
    feed_compounds=['CaCl2'],
    initial_feed_mass_balance=['Cl-']
)

cIn_span = np.geomspace(1e-3, 3.0, 61)
methods = ['debye', 'bromley', 'sit', 'pitzer']

solutions = {
    key: [
            pyequion.solve_solution( {'CaCl2': c*1e3}, sys_eq,
            activity_model_type=key)
            for c in cIn_span
        ]
    for key in methods
}
gamma_means = {key: np.array(
    [pyequion.get_mean_activity_coeff(sol, 'CaCl2')
    for sol in solutions[key]
    ])
    for key in methods
}
\end{lstlisting}

\begin{figure}[H]
	\centering
    \subfigure[][]{\includegraphics[scale=0.5]{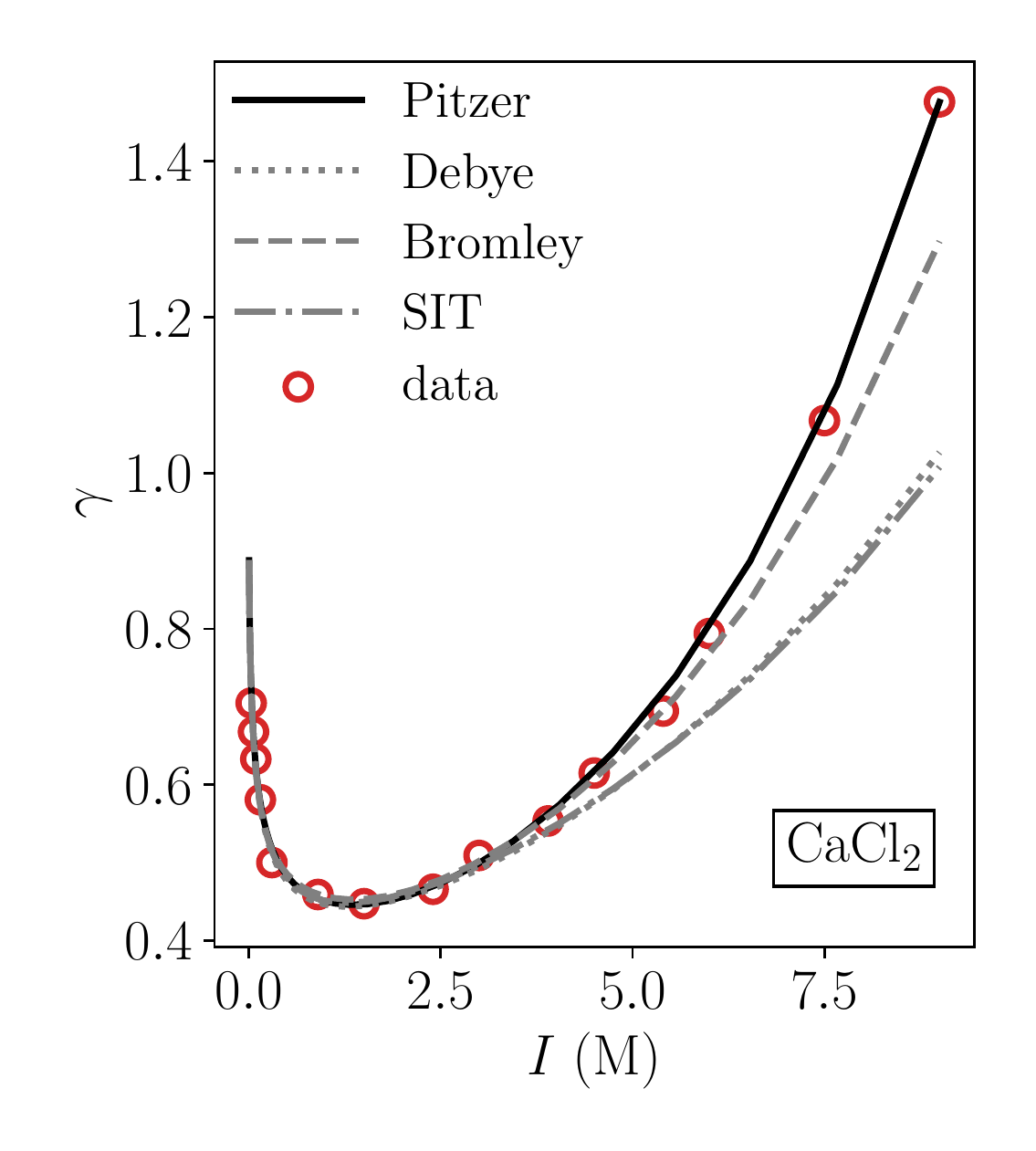}}
    \subfigure[][]{\includegraphics[scale=0.5]{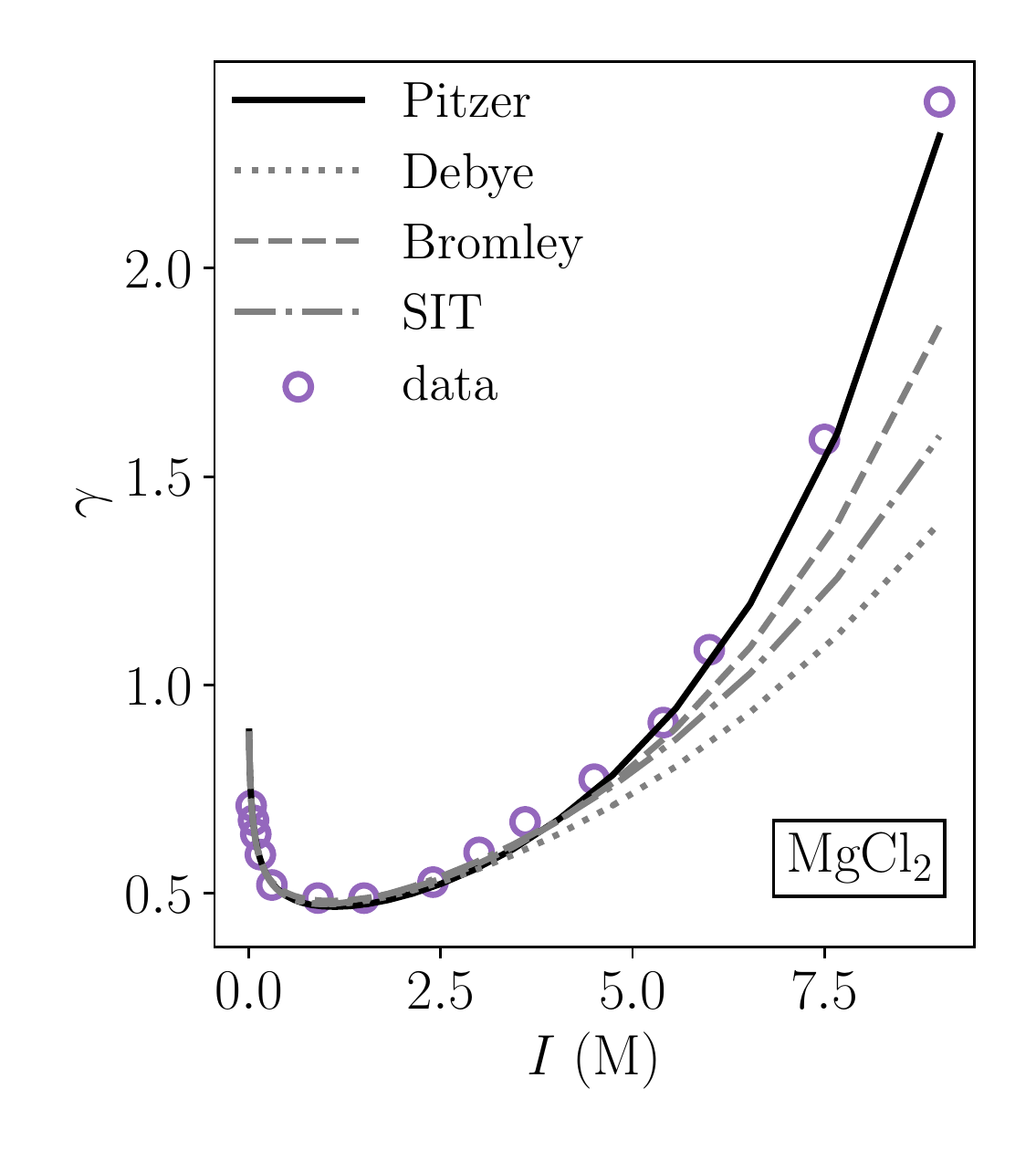}}
    \subfigure[][]{\includegraphics[scale=0.5]{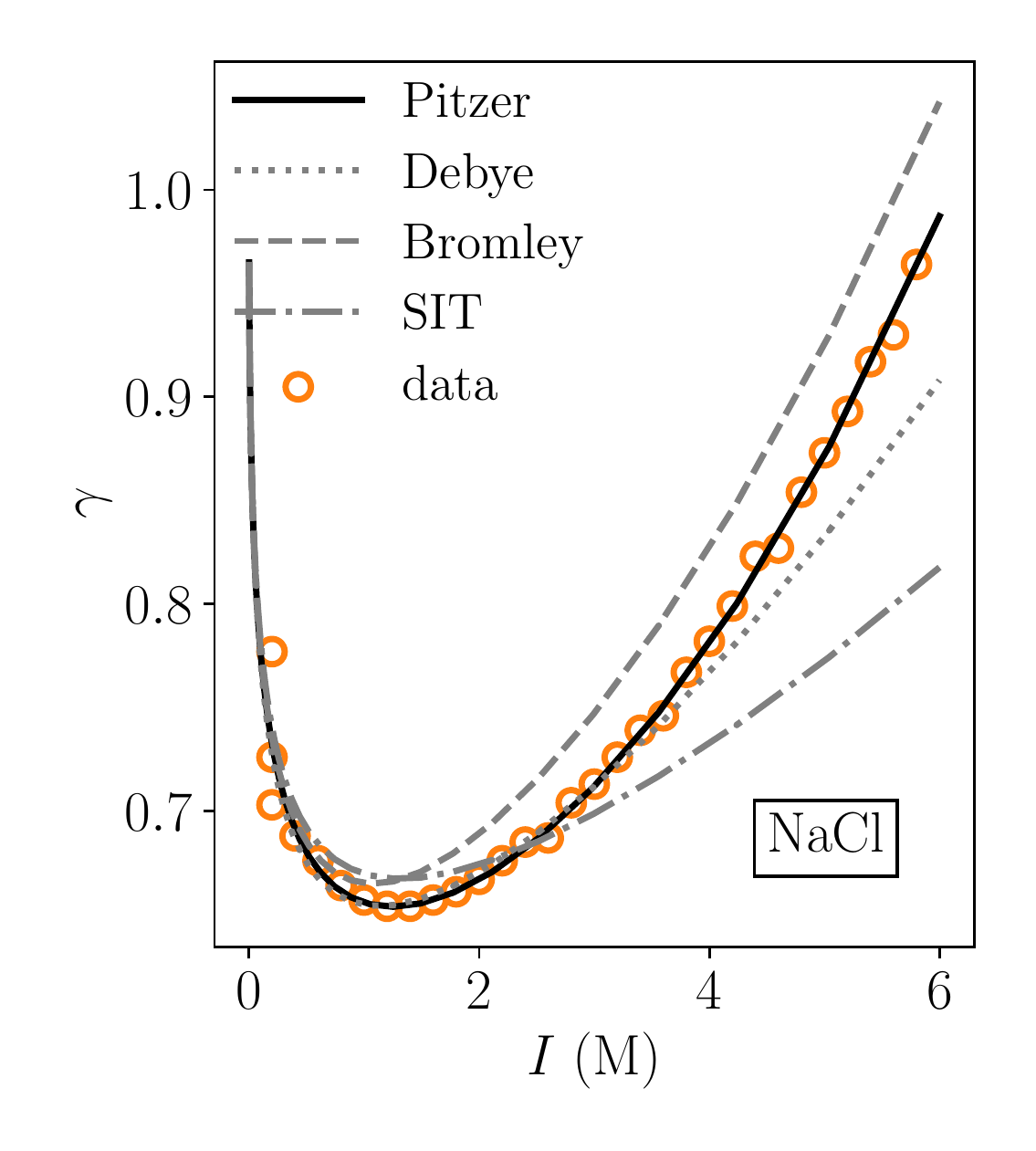}}
	\caption{Comparison of mean activity coefficient of \ce{CaCl2}, \ce{MgCl2} and \ce{NaCl} using Debye-H\"{u}ckel (\textit{B-dot}), Bromley, SIT and Pitzer models.}
	\label{fig_pt-dh-cmp}
\end{figure}

Figure \ref{fig_pt-dh-cmp} shows the comparison of the calculated mean activity coefficient with the different methods for \ce{NaCl}, \ce{CaCl2} and \ce{MgCl2} electrolytes with experimental data \citep{appelo2015principles, rodil2001individual}. The aqueous equilibrium calculation is performed and the mean activity coefficient is provided by the package using the \texttt{get{\uh}mean{\uh}activity{\uh}coeff} function. A good agreement with the experimental data is obtained when using the Pitzer model. The other models start to deviate from the experimental data for higher ionic strength. In this example, the equilibrium object is created beforehand and it is used multiple times in the calculations.

\subsection{User provided function for nonideality model}

One of the main goals of this software package is to enable easy prototyping and exploration of nonideality models. In this section, we exemplify the usage of a custom function for performing activity coefficient calculation with the Bromley method \cite{zemaitis2010handbook,bromley1973thermodynamic} using salt specific parameters.

For setting a custom function for $\log{\gamma}$ calculation, the function \texttt{solve{\uh}solution} accepts a setup and a calculation function (Listing \ref{lst-solve-bromley}). The former is used for collecting species parameters for the model and for any preprocessing, and it is executed before the nonlinear equation system. The latter is for actually calculating $\log{\gamma}$ for each species during the residual evaluation of the nonlinear system.

\begin{lstlisting}[language=Python, label=lst-solve-bromley, caption=Code sample for passing a custom activity coefficient model applied to \ce{KCl}.]
sys_eq = pyequion.create_equilibrium(['KCl'])
solution = pyequion.solve_solution({'KCl': 3e3}, sys_eq,
    setup_log_gamma_func=setup_bromley_method,
    calc_log_gamma=calc_bromley_method
)
gamma_mean = pyequion.get_mean_activity_coeff(solution, 'KCl')
\end{lstlisting}

An example of a setup function for the Bromley model is depicted in Listing \ref{lst-setup-bromley}. The model parameters can be obtained by a database file (such as JSON), but in this example, a global scoped hash table data structure is used, which contains parameters for \ce{NaCl} and \ce{KCl}. In line 1 and 2, the anions and cations are obtained, then each cation is associated with an anion parameter providing the $B_{ij}$ Bromley's interaction parameter.

\begin{lstlisting}[language=Python, label=lst-setup-bromley, caption=Code sample for a custom activity coefficient model applied to \ce{KCl}.]
bromleyDB = {
    'K+': {'Cl-': 0.024},
    'Na+': {'Cl-': 0.0574},
}
def setup_bromley_method(reaction_sys, T, db_species, c_feed):
    anions = [sp for sp in reaction_sys.species if sp.z < 0]
    cations = [sp for sp in reaction_sys.species if sp.z > 0]
    for c in cations:
        for a in anions:
            try:
                c.p_scalar[a.name] = bromleyDB[c.name][a.name]
            except KeyError:
                c.p_scalar[a.name] = 0.0
    return
\end{lstlisting}

A calculation function is presented in Listing \ref{lst-calc-bromley}. Again, the anions and cations are obtained from the list of species. The calculation is performed for each cation and anion, accordingly with the Bromley's method using parameters for the salt (not from individual ions). The function \texttt{bromley{\uh}model{\uh}ion} is not shown here, but it calculates the $\log{\gamma}_i$ of an ion $i$ giving the ionic strength $I$, interaction parameters $B_{i,j}$, species charge $z_i$ and $z_j$ and the molalities $m_j$.



\begin{lstlisting}[language=Python, label=lst-calc-bromley, caption=Code sample for a passing a custom activity coefficient model applied to \ce{KCl}.]
def calc_bromley_method(idx_ctrl, species, I, T):
    anions = [sp for sp in species if sp.z < 0]
    cations = [sp for sp in species if sp.z > 0]

    for c in cations:
        Bi_j = np.array([c.p_scalar[a.name] for a in anions])
        z_j = np.array([a.z for a in anions])
        mj = np.power(10, np.array([a.logc for a in anions]))
        loggC = bromley_model_ion(I, Bi_j, c.z, z_j, mj)
        c.set_log_gamma(loggC)

    for a in anions:
        Bi_j = np.array([c.p_scalar[a.name] for c in cations])
        z_j = np.array([c.z for c in cations])
        mj = np.power(10, np.array([c.logc for c in cations]))
        loggA = bromley_model_ion(I, Bi_j, a.z, z_j, mj)
        a.set_log_gamma(loggA)
\end{lstlisting}

\begin{figure}[H]
	\centering
    \subfigure[][]{\includegraphics[scale=0.5]{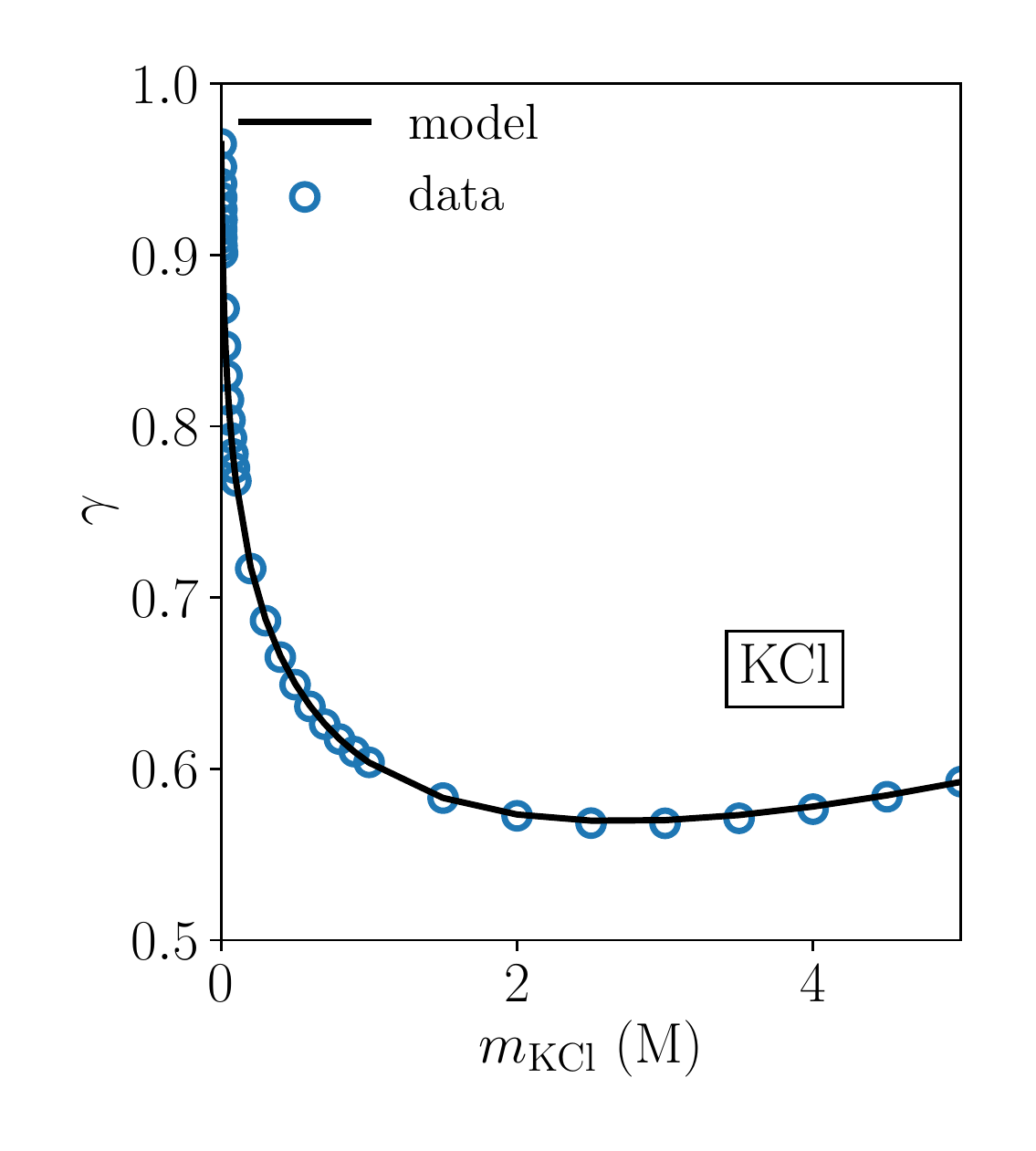}}
    \subfigure[][]{\includegraphics[scale=0.5]{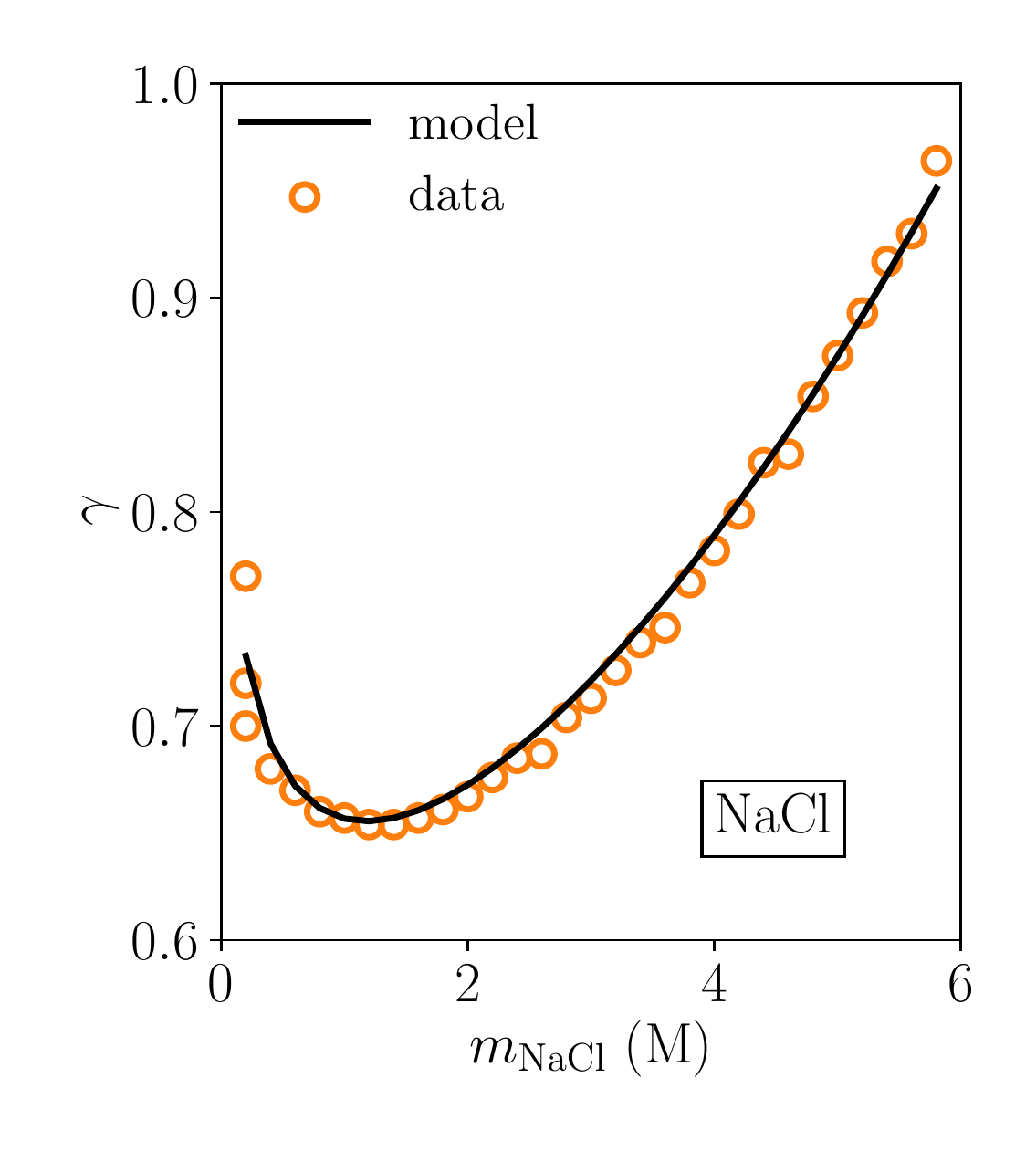}}
    \subfigure[][]{\includegraphics[scale=0.5]{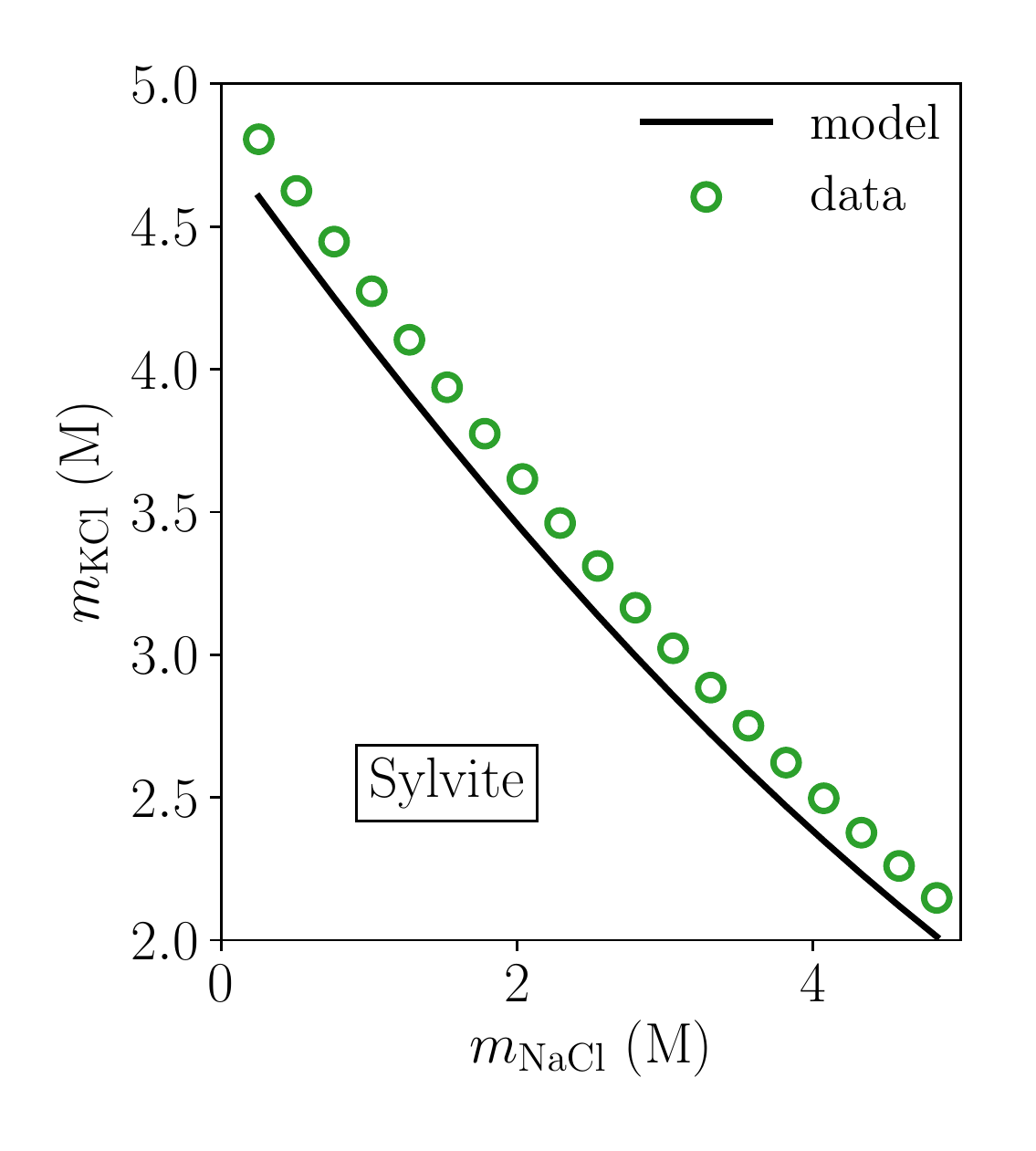}}
	\caption{Using \textit{pyequion} with a user provided nonideality model (Bromley's method with salt specific parameter) for (a) $\gamma_\text{KCl}$ and (b) $\gamma_\text{NaCl}$ single electrolyte solutions and (c) for the solubility of \ce{KCl(s)} (Sylvite) in \ce{NaCl} solution.}
	\label{fig-bromley}
\end{figure}

The Bromley's method was used for comparison with experimental data for a single electrolyte mean activity coefficient for \ce{NaCl} and \ce{KCl}. Furthermore, this model was also used for the solubility calculation of \ce{KCl} in \ce{NaCl} aqueous solution. Those results are summarized in Figure \ref{fig-bromley}.

\subsection{Embedding equilibrium calculation in dynamic simulation of crystallization}

As an illustrative example on the usage of the provided package in a simulation of a dynamic system, we show the use of \textit{pyequion} in the modelling of crystallization of calcium carbonate by the mixture of \ce{NaHCO3} with \ce{CaCl2}. The case study considers the unseeded crystallization of calcite by mixing 15 mM of \ce{NaHCO3} with 5 mM \ce{CaCl2} at a temperature of \SI{298}{\kelvin}. The crystallization model consists of a population balance solved by the Standard Method of Moments \cite{randolph_theory_1988}, which reduces the partial differential equation to a set of ordinary differential equations, and by the mass balances of total dissolved calcium and total dissolved inorganic carbon. The kinetic parameters were taken from \cite{verdoes1992determination} and are reproduced in Table \ref{tab_params_case}. The differential equations are presented by Eq. \ref{eq_odes}, where $S$ is the supersaturation, defined as $\PC{\frac{\PE{\ce{Ca^{2+}}} \PE{\ce{CO3^{2-}}}}{K_{sp}}}^{(1/2)}$ with $K_{sp}$ as the calcite solubility constant; $[\ce{Ca}]$ and $[\ce{C}]$, molal concentration of element \ce{Ca} and \ce{C}; $\rho_{c,mol}$, crystal molar density; $k_v$, shape factor; $\mu_2$, second order moment of the particle size distribution; $k_G$, growth rate constant; $g$, growth rate exponent; $k_B$, nucleation rate constant; $E_B$, activation energy for nucleation rate.

\begin{table}[H]
    \centering
    \caption{Parameters for the crystallization study using \textit{pyequion} for supersaturation calculation.}
    \begin{tabular}[]{@{}lll@{}}
        \hline
        Variable & Value & Unit\\
        \hline
        $\rho_c$ & \num{2.71e3} & \si{kg/m^3} \\
        $k_v$ & \num{1.0} & \si{-} \\
        $E_B$ & \num{12.8} & \si{-} \\
        $k_B$ & \num{1.4e18} & \si{\#/m^3.s} \\
        $g$ & \num{1.0} & \si{-} \\
        $k_G$ & \num{5.6e-10} & \si{m/s} \\
        \hline
    \end{tabular}
    \label{tab_params_case}
\end{table}

\begin{subequations}
\begin{equation}
    \dfrac{d [\ce{Ca}]}{dt} = -3 \rho_{c,mol} k_v \mu_2 \PD{k_G (S-1)^g}
\end{equation}
\begin{equation}
    \dfrac{d [\ce{C}]}{dt} = -3 \rho_{c,mol} k_v \mu_2 \PD{k_G (S-1)^g}
\end{equation}
\begin{equation}
    \dfrac{d \mu_j}{dt} =
    \begin{cases}
        k_B S \exp\PD{-\dfrac{E_B}{(\log{S})^2}} & \text{if \ } j=1\\
        j \PD{k_G (S-1)^g} \mu_{j-1}   & \text{if \ } 1<j\le 4
    \end{cases}
\end{equation}
\label{eq_odes}
\end{subequations}

\begin{figure}[H]
    \centering
    \includegraphics[scale=0.5]{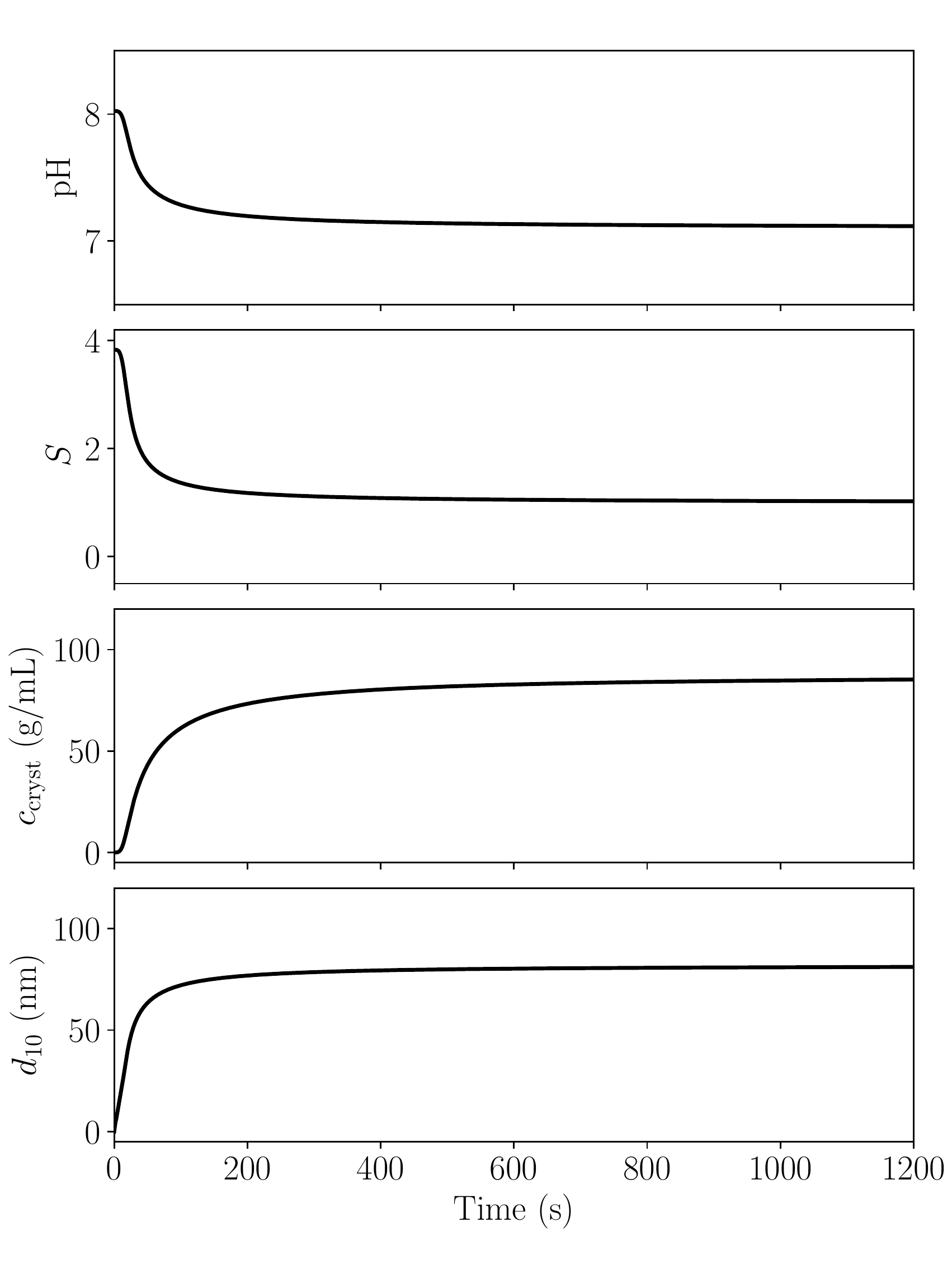}
    \caption{Dynamic profiles of pH, supersaturation $S$, crystal mass concentration  $c_\text{cryst}$ and particle mean size $d_{10}$ from the simulation of \ce{CaCO3} precipitation embedding equilibrium calculation from \textit{pyequion} exported function.}
    \label{fig:dynsim}
\end{figure}

The equilibrium is calculated from an exported function using \textit{pyequion}, as exemplified in Listing \ref{lst:export}. The equilibrium is created using the individuals ions instead of the compounds \ce{NaHCO3} and \ce{CaCl2} allowing for independent definitions of each element. The exported function is embedded in the differential algebraic equation framework \texttt{daetools} \cite{nikolic2016dae}, hence the equilibrium calculation residuals are algebraic constraints in the model. Figure \ref{fig:dynsim} presents the profiles of pH, supersaturation, suspended particles mass concentration ($c_\text{cryst}{=}\mu_3 k_v \rho_c$) and mean particle size ($d_{10}=\mu_1/\mu_0$).

\begin{lstlisting}[language=Python, label=lst:export, caption=Exporting a python function for the nonlinear residual calculation.]
sys_eq = pyequion.create_equilibrium(
    ['Na+', 'HCO3-', 'Ca++', 'Cl-'],
    fixed_elements=['Cl-'],
)
pyequion.save_res_to_file(sys_eq, './eq_nahco3_caco3.py', 'res')
\end{lstlisting}

\section{Conclusions}

Aqueous electrolyte equilibrium calculations are required in several different fields. Although there are many theoretical models available in the literature, the thermodynamic modeling of electrolyte is a controversial topic in the scientific community. In general, available software implementations of such models work well under conditions properly verified with experimental data. However, for certain operational conditions of interest, the combination of model and available parameters for calculation of reaction constants and species activity coefficients can lead to uncertainty in predictions. Many of those available softwares were written in programming languages and platforms that may be outdated, which poses a barrier for new practitioners and researchers to extend, adapt or embed into existing mathematical models.

The package \textit{pyequion} was developed in the Python language because of the easy of prototyping and the available plethora of scientific, data manipulation and visualization libraries, which makes research and learning workflows more simple. As a pure python package, \textit{pyequion} can be easily installed in different computational environments. The package can be beneficial in classroom, as the students can easily visualize the equilibrium reactions and the species for a giving mixture. A web interface was developed to allow even non-programmers to run equilibrium calculations. Because the thermodynamic models are the main part in electrolyte calculations, the software was developed in a modular manner allowing researchers to implement new models and to quickly validate the predictions for a broad range of mixtures. The equilibrium calculation can be embedded in phenomenological models using the package directly, or by exporting the nonlinear residuals to a source python function. This makes the package  suitable for inclusion in differential algebraic equation model. Therefore, \textit{pyequion} makes aqueous electrolyte equilibrium more accessible to non-experts, being suitable for both academia and technological purposes. The user should be knowledgeable of limitations and constraints when using the software, as required in any aqueous electrolyte thermodynamic prediction.

Because of the relevance of aqueous electrolyte equilibrium in basic science and engineering applications, efforts on software development that facilitates this type of calculation are important. In this work, we report a pure Python package for equilibrium calculation of aqueous electrolyte systems. The package can determine the reactions and species in a given system, and it calculates equilibrium concentrations, which makes it a reliable and flexible tool for students, researchers and practitioners in technological applications.

\section{Declaration of Competing Interest}

We wish to confirm that there are no known conflicts of interest associated with this publication and there has been no significant financial support for this work that could have influenced its outcome.

\section*{Acknowledgements}

This research was carried out in association with the ongoing R\&D project registered as ANP n\textdegree \ 21609-3, ``Modelagem preditiva de incrusta\c{c}\~{o}es inorg\^{a}nicas em escoamentos multif\'{a}sicos em tubula\c{c}\~{o}es e equipamentos para campos do pr\'{e}-sal.`` (UFRJ/Shell Brazil/ANP), sponsored by Shell Brazil under the ANP R\&D levy as "Compromisso de Investimentos com Pesquisa e Desenvolvimento". We gratefully acknowledge support of Shell Brazil and ANP.





\section*{References}

\bibliographystyle{elsarticle-num}
\bibliography{manuscript}







\end{document}